\documentclass[prl,aps,twocolumn,showpacs,floatfix,longbibliography]{revtex4-2}
 
\usepackage[applemac]{inputenc}
\usepackage{stmaryrd}
\usepackage{mathtools} 
\usepackage{stmaryrd}
\usepackage{stackengine}
\stackMath

\newcommand {\Tr}{{\rm Tr}\,}

\newcommand {\Sp}{{\rm Sp}\,}
\newcommand{\comm}[1]{}
\usepackage{amsmath}
\usepackage{amsfonts}

\usepackage{amssymb}
\usepackage{graphicx}%
\usepackage{tabularx}
\usepackage{tikz}

\newcommand{\op}[1]{\hat{#1}}         
\newcommand{\bvec}[1]{\bm{#1}}        
\newcommand{\gam}[1]{\gamma_{#1}}

\begin{document}

\title{ Relativistic Lindblad  description of the  electron's radiative dynamics}

\date{\today}

\author{Andre G. Campos}
\email{agontijo@mpi-hd.mpg.de}
\author{Karen Z. Hatsagortsyan}
\author{Christoph H. Keitel}
\affiliation{Max Planck Institute for Nuclear Physics, Heidelberg 69117, Germany}

\begin{abstract}

An effective model for describing the relativistic quantum dynamics of a radiating electron is developed via a relativistic generalization of the Lindblad master equation. By incorporating both radiation reaction and vacuum fluctuations into the Dirac equation within an open quantum system framework, our approach captures the Zitterbewegung of the electron,  ensuing noncommutativity of its  effective spatial coordinates,  and provides the quantum analogue of the Landau-Lifshitz (LL) classical equation of motion with radiation reaction. We develop the corresponding phase-space representation via the relativistic Wigner function and derive the semiclassical limit through a Foldy-Wouthuysen transformation. The latter elucidates the signature of quantum vacuum fluctuations in the LL equation, and shows its relationship with the corrected Sokolov equation. 
Our results offer a robust framework for investigating quantum radiation reaction effects in ultrastrong laser fields.

\end{abstract}

\maketitle

\textit{Introduction.} Recent advancement of ultrastrong laser techniques \cite{Yoon_2021,Danson_2019} and experiments in ultrastrong laser facilities  \cite{Cole_2018,Poder_2018} revived theoretical interest to  the problem of the electron radiation reaction (RR) \cite{di2012extremely}. In external fields, an electron radiates and loses energy and momentum, which in classical theory is accounted for by introducing the RR force in the electron equation of motion. The latter is derived excluding radiation variables from the consideration which leads to the Lorentz-Abraham-Dirac (LAD) equation \cite{Dirac_1938}, and to its iterated versions such as Landau-Lifshitz (LL) \cite{landau2013classical},  or Sokolov equations~\cite{sokolov2009renormalization,sokolov2009dynamics}, or similar alternative equations \cite{Eliezer_1948,Mo_1971,Ford_1993,Seto_2011}. In the quantum regime, the accurate treatment of RR via multiple photon emissions is cumbersome \cite{di2010quantum}, and usually QED Monte Carlo simulations \cite{Duclous_2011,Ridgers_2014,Gonoskov_2015} are applied employing a heuristic approach in which the electron dynamics is accounted for classically and the photon emissions by QED probabilities in the locally constant field approximation (LCFA). An intuitive kinetic equation with a similar approach has been put forward by Baier and Katkov \cite{Baier_b_1994},  applied also for the laser driven setup  \cite{neitz2013stochasticity}. Derivation of the kinetic equation from first principles treating quantum RR rigorously is shown recently in Ref.~\cite{Fauth_2021}, which however leads to a system of  equations unwieldy for applications. The QED kinetic equations in strong fields   \cite{Kluger_1998,Schmidt_1998,Roberts_2002,Gies_2005,Hebenstreit_2008,Hebenstreit_2009,Hebenstreit_2010,
Blinne_2014,Edwards_2025} are developed for treating electron-positron creation and its back reaction and do not include the emitted photon variables.

When excluding radiation variables from the consideration, the radiating system of electrons quantum mechanically represents an open system. An open quantum system is rigorously described by a density matrix obeying a Lindblad master equation  \cite{lindblad1976generators,gorini1976completely,breuer2002theory,gardiner2004quantum,
manzano2020short}. The dynamics of such an equation obeys the uncertainty principle, avoids negative probabilities, and conserves the total probability.  Lindblad master equations were employed to describe 
quantum Brownian motion \cite{boyanovsky2015effective}, quantum relativistic diffusion \cite{arnault2020quantum}, quantum friction \cite{bondar2016wigner}, the inelastic loss of ultracold atoms \cite{braaten2017lindblad},  postquantum classical gravity \cite{Oppenheim_2023}, entropic gravity \cite{sung2023decoherence}, \textit{etc}. However, the extension of the Lindblad model for a dissipating system  into the relativistic domain is problematic, as loosing energy leads to unphysical fall of electrons down the negative energy ladder.

In standard derivation of the Lindblad equation, usually one starts with the full system including the environment, and further the environment variables are excluded treating it as  perturbation, and the Lindblad relaxation operators are derived by corresponding perturbative expansion. However, this approach is exceedingly complex, burdened by multiple approximations and significant limitations.
A more efficient method of operational dynamical modeling (ODM) has been proposed in Ref.~\cite{bondar2016wigner}  for the nonrelativistic problems. In ODM, Lindblad operators are heuristically constructed in a such form to lead to the Ehrenfest equation in a physically relevant form.

The Ehrenfest equation with appropriate modifications  was applied for the description of QED effects as an interaction with an environment. Thus, Welton~\cite{Welton1948}  associated the electromagnetic shift of energy levels in atoms with oscillations in the position of an electron interacting with the zero-point fluctuations of electromagnetic fields,  \textit{vacuum fluctuations} (VF), and derived within this picture the Lamb shift and the anomalous magnetic moment of the electron. The VF was shown to induce \textit{Zitterbewegung} \cite{koba1949semi}  modifying the Ehrenfest relation for the effective position operator \cite{tani1951connection}.

In this Letter, we develop a relativistic framework for dissipative open Dirac systems, extending nonrelativistic ODM to include RR as an environmental effect within the Lindblad master equation, bringing about a quantum analogue of the classical LAD approach. We remedy the inherent instability of relativistic dissipative systems by introducing a new Lindblad term describing VF, employing ideas of Refs.~\cite{Welton1948,koba1949semi,tani1951connection} within ODM. The VF term not only effectively suppresses runaway solutions by rendering transitions to negative energy states virtual, but also generates a noncommutative geometry for  the effective spatial coordinates, linked to the electron Zitterbewegung and leaving VF imprints on the classical equations of motion with RR. We construct the phase space representation of the obtained Lindblad equation via the relativistic Wigner function, and derive the semiclassical limit. Finally, we demonstrate the connection between the modified Ehrenfest equation and the classical LL and  Sokolov equations (the latter with the corrected VF contribution) and discuss their relationship.

\textit{Lindblad Master Equation.} We begin with the density matrix $\op{P}(t)$ for an open quantum system in the Schr\"odinger picture, which evolves according to the Lindblad equation \cite{manzano2020short} (see Sec.~I of the Supplemental Material (SM)~\cite{SupplementalM}): 
\comm{
For an open quantum system, the density matrix $\op{P}(t)$ in the Schr\"odinger picture evolves according to  to the Lindblad equation \cite{manzano2020short}): 
\begin{multline}
i\,\frac{\partial \op{P}(t)}{\partial t} = -\bigl[\op{P}(t),\op{H}(t)\bigr]
+ i\sum_\lambda\sum_{\eta=\pm1}\int_{-\infty}^{\infty}d\tau\, \\[1mm]
\int\!\frac{d^3\mathbf k}{(2\pi)^{3/2}\omega}\Bigl(
\op{B}_{\eta,\lambda}\Bigl(t_1,k\Bigr)\op{P}(t)
\op{B}_{\eta,\lambda}^\dagger\Bigl(t_2,k\Bigr)\\[1mm]
-\frac{1}{2}\Bigl\{
\op{B}_{\eta,\lambda}^\dagger\Bigl(t_2,k\Bigr)
\op{B}_{\eta,\lambda}\Bigl(t_1,k\Bigr),
\op{P}(t)
\Bigr\}
\Bigr),
\label{eq:lindblad_master_full_pre}
\end{multline}
where  $\op{P}\gam{0}=|\psi\rangle\langle\psi|\gam{0}$ with a Dirac spinor $|\psi\rangle$, the hat indicates a self-adjoint operator, $\{.,.\}$ and $[.,.]$  stand for the anticommutator and commutator, respectively and $t_1=t-\tau/2$, $t_2=t+\tau/2$. Relativistic units with $\hbar=c=1$ are used throughout. In the interaction picture $\op{P}_I(t)=e^{i\hat{H}t}\op{P}(t)e^{-i\hat{H}t}$ we have
}
\begin{multline}
\frac{\partial \op{P}(t)}{\partial t} = i\bigl[\op{P}(t),\op{H}(t)\bigr]+
\sum_\lambda\sum_{\eta=\pm1}\int_{-\infty}^{\infty}d\tau\, 
\int\!\frac{d^3\mathbf k}{(2\pi)^{3/2}\omega}\\[1mm]
\Bigl(
\op{B}_{\eta,\lambda}\Bigl(-\tau/2,k\Bigr)\op{P}(t)
\op{B}_{\eta,\lambda}^\dagger\Bigl(\tau/2,k\Bigr)\\[1mm]
-\frac{1}{2}\Bigl\{
\op{B}_{\eta,\lambda}^\dagger\Bigl(\tau/2,k\Bigr)
\op{B}_{\eta,\lambda}\Bigl(-\tau/2,k\Bigr),
\op{P}(t)
\Bigr\}
\Bigr).
\label{eq:lindblad_master_full}
\end{multline}
where $\op{P}\gam{0}=|\psi\rangle\langle\psi|\gam{0}$ with a Dirac spinor $|\psi\rangle$, the hat indicates a self-adjoint operator, $H$ is the Hamiltonian of the system, $\{.,.\}$ and $[.,.]$  stand for the anticommutator and commutator, respectively and $t_1=t-\tau/2$, $t_2=t+\tau/2$. Relativistic units with $\hbar=c=1$ are used throughout.  
The Lindblad equation ensures complete positivity and conservation of probability, and forms the starting point for our analysis. 
Employing the ODM, we will construct the  dissipating operators $\op{B}_{\eta,\lambda}(t,k)$  to fit the desired Ehrenfest equation with RR. 

It is known that the RR induces a friction term into the Ehrenfest equation, which is proportional to the velocity in the quasiclassical regime, see e.g. \cite{sokolov2009renormalization}. Therefore, the Ehrenfest equation obtained from the Lindblad model is expected  to have the following form in this limit:
 \begin{eqnarray}\label{EhrenfestFriction}
\frac{d\langle\hat{\boldsymbol{x}}^k\rangle}{dt}=\langle\gamma^0\gamma^k\rangle, \,\, \frac{d\langle\hat{\boldsymbol{\pi}}_k\rangle}{dt}=\langle F_{k\nu}\gamma^0\gamma^\nu\rangle -\sigma  m\langle\gamma_k\gamma_0\rangle , 
\end{eqnarray}
where $\langle\hat{\boldsymbol{x}}^k\rangle$ and $\langle\hat{\boldsymbol{\pi}}_k\rangle$ are the expectation values of the coordinate and  kinetic momentum, respectively, $\gamma_k$ are the Dirac matrices, $F_{k\nu}$ is the field tensor of the background field, $m$  the electron mass, and $\sigma$ a still unknown RR parameter which will be deduced comparing our final results with those of quasiclassical nonlinear QED.

Taking into account that RR emerges due to photon emissions, we define the Lindblad operator that encodes RR as
\begin{eqnarray}
		&\hat B_{\eta,\lambda}(t,k)
				=U(t)\frac{\sigma^{1/2}e^{i\eta\,\omega t}}{\sqrt{2}}\;
		e^{-i\eta\,\mathbf k\cdot\hat{\mathbf x}}\;
		\boldsymbol\alpha\cdot\mathbf e_\lambda U(t)^\dagger
			\label{eq:Hjump}
\end{eqnarray}
where $\eta=\pm$ corresponds to the photon emission (absorption) with the photon 4-vector $k=(\omega,\mathbf{k})$, and the  polarization $\mathbf e_\lambda$, $\boldsymbol{\alpha}$ are the Dirac matrices, and $U(t)$ is the time evolution operator. Our aim within ODM is to show that the Lindblad operator  [Eq.~(\ref{eq:Hjump})] in the quasiclassical limit leads to the damping term in the Ehrenfest equation proportional to the velocity.

Thus, we consider the case 
when the electron dynamics in the external field is quasiclassical between the photon emissions.
We will show that the dissipating term in the Ehrenfest equation is expressed via the photon emission probabilities in strong fields of the Baier-Katkov QED operator method~\cite{Baier_b_1994,landau1982course,schwinger1951theory}. For the quasiclassical dynamics, 
the effects of negative energy states (NES) should be excluded in the dissipating operators by transforming (\ref{eq:Hjump}) to the Foldy-Wouthuysen (FW) representation via the unitary operator $U_{FW}(\hat{\boldsymbol{\pi}}(t))$, and  applying the projector $P_+=(\mathbf{1}+\gamma_0)/2$ on both sides, see End Matter. The final result is
\begin{eqnarray}
\label{eq:HjumpFW}
&\tilde B_{\eta,\lambda}(t,k)=\sqrt{\frac{\sigma}{2}}e^{i\eta\,\omega t}
e^{-i\eta\,\mathbf k\cdot\hat{\mathbf x}(t)}R(t), 
\end{eqnarray}
with $R(t)= P_+U_{FW}(\hat{\boldsymbol{\pi}}'(t))\bigl(\boldsymbol\alpha\!\cdot\!\mathbf e_\lambda\bigr)U_{FW}(\hat{\boldsymbol{\pi}}(t))^\dagger P_+=$\\$ =\mathbf{e}_\lambda\cdot\mathbf{A}(t)+i\boldsymbol{\sigma}\cdot\mathbf{\mathcal{B}}(t)$, and 
\begin{align*}
\mathbf{A}(t)&=\frac{1}{2\sqrt{\boldsymbol{\pi}_0\boldsymbol{\pi}'_0}}\left(\hat{\boldsymbol{\pi}}(t)\sqrt{\frac{\boldsymbol{\pi}'_0+m}{\boldsymbol{\pi}_0+m}}+\hat{\boldsymbol{\pi}}'(t)\sqrt{\frac{\boldsymbol{\pi}_0+m}{\boldsymbol{\pi}'_0+m}}\right),\\
\mathbf{\mathcal{B}}(t)&=\frac{\mathbf{e}_\lambda}{2\sqrt{\boldsymbol{\pi}_0\boldsymbol{\pi}'_0}}\times\left(\hat{\boldsymbol{\pi}}(t)\sqrt{\frac{\boldsymbol{\pi}'_0+m}{\boldsymbol{\pi}_0+m}}-\hat{\boldsymbol{\pi}}'(t)\sqrt{\frac{\boldsymbol{\pi}_0+m}{\boldsymbol{\pi}'_0+m}}\right).
\end{align*}
Here $\boldsymbol{\pi}'_0=\boldsymbol{\pi}_0-\eta\omega/2$ and $\hat{\boldsymbol{\pi}}'(t)=\hat{\boldsymbol{\pi}}(t)-\eta\mathbf{k}/2$, see  Sec.~IV of SM~\cite{SupplementalM}.

To evaluate the dissipating term in the momentum Ehrenfest equation~(\ref{EhrenfestFriction}), we calculate the expectation value $\langle\hat{O}\rangle=\frac{1}{4}\Tr [\hat{P}\hat{O} ]$ of a general observable $\hat{\mathcal{O}}$ related to the Ehrenfest relation which stems from the Lindbad dissipating operator of Eq.~(\ref{eq:Hjump}): 
\begin{align}\label{EhrenfestRR}
\mathcal{L}_{RR}^\dagger(\hat{\mathcal{O}})&=\int d\tau\sum_{\eta=\pm1}\int\frac{d^3\boldsymbol{k}\sigma }{2(2\pi)^{3}\omega}
\sum_\lambda\mathcal{T},
\end{align}
where { $t_1=t-\tau/2$, and $t_2=t+\tau/2$, and   
\begin{align*}
\mathcal{T}&=e^{-i\eta\omega\tau}R_2^\dagger e^{i\eta\mathbf{k}\cdot\hat{\mathbf{x}}(t_2)}\hat{\mathcal{O}}e^{-i\eta\mathbf{k}\cdot\hat{\mathbf{x}}(t_1)}R_1\\
&-\frac{e^{-i\eta\omega\tau}}{2}\{\hat{\mathcal{O}},R_2^\dagger e^{i\eta\mathbf{k}\cdot\hat{\mathbf{x}}(t_2)}e^{-i\eta\mathbf{k}\cdot\hat{\mathbf{x}}(t_1)}R_1\},
\end{align*}
with $R_i=R(t_i)$. For $\hat{\mathcal{O}}=\hat{\boldsymbol{\pi}}$, we have
$
\mathcal{T}=-2\eta \mathbf{k}e^{-i\eta\omega\tau}R_2^\dagger R_1e^{i\eta\mathbf{k}\cdot\hat{\mathbf{x}}(t_2)}e^{-i\eta\mathbf{k}\cdot\hat{\mathbf{x}}(t_1)}.
$
After tracing out the spin degrees of freedom gives via $\Tr[R_2^\dagger R_1]/2=\mathbf{A}_2^\dagger\cdot\mathbf{A}_1+\mathcal{B}_2^\dagger\cdot\mathcal{B}_1$, and averaging over the photon polarization, we obtain in the ultrarelativistic regime $\boldsymbol{\pi}_0\gg m$:
\begin{eqnarray}
\overline{\mathcal{L}}_{RR}^\dagger(\hat{\boldsymbol{\pi} })&=& 
-\frac{\eta\mathbf{k}}{(\boldsymbol{\pi}'_0)^2}\left(((\boldsymbol{\pi}'_0)^2+\boldsymbol{\pi}_0^2)(\mathbf{v}_2\cdot\mathbf{v}_1-1)+\frac{\omega^2}{\gamma^2}\right)L,\nonumber\\
 L&=& \exp\Big[i\eta\frac{\boldsymbol{\pi}_0}{\boldsymbol{\pi}_0'}(-\omega\tau+\mathbf{k}\cdot(\mathbf{x}(t_2)-\mathbf{x}(t_1))\Big],
\label{LLL}
\end{eqnarray}
where $\mathbf{v}_i=\boldsymbol{\pi}_i/\boldsymbol{\pi}_0$, and $\overline{\mathcal{L}}_{RR}^\dagger(\hat{\boldsymbol{\pi} })\equiv \frac{1}{2}\sum_\lambda\Tr[\mathcal{T}]$. In the latter the radiation probability is expressed via the electron classical trajectory as in the Baier-Katkov theory.  A more general expression for (\ref{LLL}) is given in Sec.~III of SM~\cite{SupplementalM}.
In the limit of small recoil the Lindbad dissipating operator induces the friction term in the Ehrenfest  equation proportional to velocity:
$\overline{\mathcal{L}}_{RR}^\dagger(\hat{\mathbf{p}})\approx-\frac{\sigma  \gamma ^4 |\ddot{\mathbf{x}}|^2}{2\pi  }\dot{\mathbf{x}}$, see End Matter, fulfilling our assumption and justifying the choice of Eq.~(\ref{eq:Hjump}) for the Lindbad dissipating operators.  However, the relativistic Lindbad equation (\ref{eq:lindblad_master_full}) with the sole friction term is unstable, decaying infinitely to NES.

\textit{Vacuum fluctuations.} In order to solve the stability problem, we will introduce the Lindblad operator that effectively prevents runway solutions. We introduce a set of Lindblad operators $V_\pm$ that implement 
negative energy state (NES) electron transition into a positive energy state (PES), and  vice versa,  modeling VF, see End Matter, and add VF term in Eq.~(\ref{eq:lindblad_master_full}):
\begin{eqnarray}
\mathcal{L}_{VF}(\hat{P})=i\sum_{s=\pm1} \sigma_s\Biggl(
\op{V}_{s} \op{P}(t)
\op{V}_{s}^\dagger  -\frac{1}{2}\Bigl\{
\op{V}_{s}^\dagger 
\op{V}_{s} 
\op{P}(t)
\Bigr\}
\Biggr). 
\end{eqnarray}
 The VF not only stabilize the Lindbad model, but also modifies the Ehrenfest relation for the position operator. Let us illustrate the latter VF effect for a free electron:  
\begin{eqnarray}\label{EhrenfestPFields}
&\frac{d}{dt}\langle\hat{\boldsymbol{x}}_k\rangle
=\Tr\left(i[\hat{H}_0,\hat{\boldsymbol{x}}_k]+\hat{\boldsymbol{x}}_k \left(\mathcal{L}_{RR}(\hat{P})+ \mathcal{L}_{VF}(\hat{P})\right)\right)
\end{eqnarray}
with the free Hamiltonian  $H_0$. The first term is trivial and the second is zero, while the third term is the desired contribution of VF. For  the free electron  $ \hat{V}_{\pm}=\pm if\gamma_5\gamma_0\hat{\mathcal{P}}_\mp$, see Eq.~(\ref{Creation_annihilation}) in End Matter, with the projector operators for PES (NES) $\hat{\mathcal{P}}_\pm $.
Thus,  the VF modify the Ehrenfest  equation for the position  
\begin{align}\label{EhrenfestXfields}
\frac{d\langle\hat{\boldsymbol{x}}^k\rangle}{dt}&=\langle\gamma^0\gamma^k\rangle-(\sigma_++\sigma_-)\frac{\lambdabar}{2}\left\langle i\gamma^k+\varepsilon^{ijk}\lambdabar\hat{\boldsymbol{p}}_i\boldsymbol{\Sigma}_j\right\rangle,
\end{align}
where $\boldsymbol{\Sigma}_j=-i\gamma_5\gamma_j\gamma_0$ is the spin operator, $\varepsilon_{ijk}$ is the Levi-Civita tensor, and $\lambdabar$ is the Compton wavelength. 
The last term  in Eq.~(\ref{EhrenfestXfields}) is the VF correction, which can be interpreted via the eminent analysis of Refs.~\cite{Welton1948,koba1949semi,tani1951connection,pryce1948mass,barut1981Zitterbewegung}.
For an ensemble of noninteracting electrons, 
a ``centre of mass" can be defined \cite{pryce1948mass}, and the VF induced term  is proportional to Pryce's centre of mass coordinate $\mathbf{x}-\mathbf{q}$,
relative to which  \textit{Zitterbewegung} occurs~\cite{barut1981Zitterbewegung}, which  
is determined by the property of noncommutativity of space coordinates   $[\boldsymbol{x}_i-\boldsymbol{q}_i,\boldsymbol{x}_j-\boldsymbol{q}_j]\sim \lambdabar/2 $~\cite{Welton1948}.   The \textit{Zitterbewegung} originated from  VF  gives rise to an additional term in the electron's current, besides implementing the stability of the relativistic Lindbad equation.

\textit{Phase Space Representation.} As an alternative to the solution of the Lindbad equation, we derive the corresponding kinetic equation for the  Wigner function, which also facilitates the connection between the quantum and classical regimes.
We introduce the following covariant Wigner function:
\begin{align}\label{NoNCovWF}
W(x,\boldsymbol{\pi})&=\int d^3\mathbf{y}\exp\left[\frac{-i\mathbf{p}\cdot \mathbf{y}}{\hbar}-\frac{ie}{\hbar c}\int_{\mathbf{x}-\frac{\mathbf{y}}{2}}^{\mathbf{x}+\frac{\mathbf{y}}{2}}d\boldsymbol{z}^k \boldsymbol{A}_k(\mathbf{z}) \right]\nonumber\\
&\times P\left(x+\frac{y}{2},x-\frac{y}{2}\right).
\end{align}
where $x=(x^0,\mathbf{x})$ and $P(x+\tfrac{y}{2},x-\tfrac{y}{2})$ is the density matrix in coordinate space. The Wigner function acts as a quasi-probability distribution whose marginals yield the probability densities in coordinate and momentum space.
The time evolution of the Wigner function is governed by the Lindblad--Wigner--Dirac equation:
\begin{align}
	&i\frac{\partial}{\partial t}W=-[W,H]+i\mathcal{L}_{RR}(W)+i\mathcal{L}_{VF}(W), \label{MasterWigner}\\
	&\mathcal{L}_{RR}(W)=\frac{1}{2}\sum_{\eta=\pm1}\int_{-\infty}^\infty d\tau\int\frac{d^3\mathbf{k}}{(2\pi)^3\omega}\Big[\Big(2B_{\eta,I}(t_2,k)\star\nonumber\\
	& W\star B_{\eta}(t_1,k)^\dagger
	 -W\star B_{\eta}(t_1,k)^\dagger\star B_{\eta}(t_2,k)\nonumber\\
	 &- B_{\eta}(t_1,k)^\dagger\star B_{\eta}(t_2,k)\star W\Big)\Big)\Big],\label{RRW}\\
	&\mathcal{L}_{VF}(W)=\frac{1}{2}\sum_{s=\pm}\sigma_{s}\Big(2V_{s}\star W\star V_{s}^\dagger-W\star V_{s}^\dagger\star V_{s}\nonumber\\
	&-V_{s}^\dagger\star V_{s}\star W\Big),\label{VFW}\\
	&\star = \exp \frac{i\hbar}{2} \left(\sum_{j=1}^3\left[
		\overleftarrow{\frac{\partial}{\partial \boldsymbol{x}^j}} \overrightarrow{\frac{\partial}{\partial \boldsymbol{\pi}^j}} -
		\overleftarrow{\frac{\partial}{\partial \boldsymbol{\pi}^j}} \overrightarrow{\frac{\partial}{\partial \boldsymbol{x}^j}}\right]
	 \right) \exp\left(i\sum_{n=1}^\infty\hbar^{n}\mathbf{L}_n \right).\label{EqMoyalStar}
\end{align}
were the gauge invariant star product is defined \cite{mueller1999product}. In Eq.~(\ref{EqMoyalStar}) the commutation of the two exponential terms is used. The upper arrows indicate the direction in which the derivatives act. In the case of the time-independent  external fields, we have  
\begin{align*}
&\mathbf{L}_n=\left(\frac{i}{2}\right)^{n+1}\frac{\epsilon^{jlr}}{(n+1)^2n!}\sum_{i_1,...,i_{n-1}=1}^3\left(\frac{\partial^{n-1}}{\partial \boldsymbol{x}_{i_1}...\partial  \boldsymbol{x}_{i_{n-1}}}e\boldsymbol{B}_r\right)\\
&\times\overleftarrow{\frac{\partial}{\partial \boldsymbol{\pi}_j}} \overrightarrow{\frac{\partial}{\partial \boldsymbol{\pi}_l}}\sum_{p=1}^n\binom{n+1}{p}g(n,p)\overleftarrow{\frac{\partial}{\partial \boldsymbol{\pi}_{i_1}}}...\overleftarrow{\frac{\partial}{\partial \boldsymbol{\pi}_{i_{p-1}}}}
\overrightarrow{\frac{\partial}{\partial \boldsymbol{\pi}_{i_p}}}...\overrightarrow{\frac{\partial}{\partial \boldsymbol{\pi}_{i_{n-1}}}},
\end{align*}
with $e\boldsymbol{B}_r=\epsilon_{rjk}\partial^j e\boldsymbol{A}^k$ and $g(n,p)=[(1-(-1)^p)(n+1)-(1-(-1)^{n+1})p]$. 
Most of the information of  the relativistic Wigner function is contained in $W^0(x , \boldsymbol{\pi})  \equiv \Sp [W(x , \boldsymbol{\pi})]/4$ \cite{campos2014violation}, where $\Sp[\cdot]$ is the trace  over the spin variables, because it determines the probability density in both coordinate and momentum space: $\int_{-\infty}^\infty  W^0(x ,\boldsymbol{\pi}) d^3\boldsymbol{\pi} =\rho(x)$,  $\int_{-\infty}^\infty  W^0(x ,\boldsymbol{\pi}) d^3\mathbf{x} =  \tilde{\rho}(t,\boldsymbol{\pi})$. 

We can simplify the Wigner kinetic equation in the case of the quasiclassical dynamics in the background fields, following the  Baier-Katkov concept.
 Taking into account that in the Moyal product the gradient terms are proportional to the quasiclasical parameters of $\lambda_{dB}/\ell$, with the de-Broglie wavelength $\lambda_{dB}$, and the typical length  $\ell$ of the external field, and neglecting them in the quasiclassical approximation,  we obtain 
in ultrarelativistic regime $\gamma \gg 1$, see End Matter:
\begin{eqnarray}
\label{LRRf}
&&\mathcal{L}_{RR}(W)=\frac{\sigma}{2}\sum_{\eta=\pm1}\int_{-\infty}^\infty d\tau\int\frac{d^3\mathbf{k}}{(2\pi)^3\omega}\mathcal{L}_{RR,\mathbf{k}}(W),\\
&&\mathcal{L}_{RR,\mathbf{k} }(W)=-\frac{P_{ij}(\mathbf n)}{2}\left((\alpha_1^i  L(\tau)\alpha_2^j) \overleftarrow{W}+\overrightarrow{W}(\alpha_1^i L(\tau)\alpha_2^j)\right)\nonumber\\
&+& P_{ij}(\mathbf n)\big(\alpha_{2}^i L(\tau/2) \overleftarrow{W}\Big(t,\mathbf{x}+\frac{i\hbar\nabla_{\boldsymbol{\pi}}}{2},\boldsymbol{\pi}'\Big)\big)  \overleftarrow{L(-\tau/2)^\dagger\alpha_{1}^j},\nonumber
\end{eqnarray}
with $L(\tau)$ of Eq.~(\ref{LLL}) and $P_{ij}(\mathbf n)\equiv\sum_\lambda e_{\lambda i}^\ast e_{\lambda j}=\delta_{ij}-n_i n_j$. In the same approximation VF term reads
\begin{eqnarray}
\label{LWFf}
\mathcal{L}_{VF}(W)&=&\sigma_{s}\Big(\left(\gamma_5\gamma_0I_s(\boldsymbol{\pi}) \overleftarrow{W}\left(t,\mathbf{x}+\frac{i\hbar\nabla_{\boldsymbol{\pi}}}{2},\boldsymbol{\pi}\right)\right) \\
&\times & \overleftarrow{I_s}\left(\boldsymbol{\pi}-\frac{i\hbar\nabla_{\mathbf{x}}}{2}\right)\gamma_5\gamma_0\nonumber\\
&-&\frac{1}{2}G(\mathbf{x},\boldsymbol{\pi})  \overleftarrow{W}\left(t,\mathbf{x}+\frac{i\hbar\nabla_{\boldsymbol{\pi}}}{2},\boldsymbol{\pi}-\frac{i\hbar\nabla_{\mathbf{x}}}{2}\right)\nonumber\\
&-&\frac{1}{2}\overrightarrow{W}\left(t,\mathbf{x}-\frac{i\hbar\nabla_{\boldsymbol{\pi}}}{2},\boldsymbol{\pi}+\frac{i\hbar\nabla_{\mathbf{x}}}{2}\right) G(\mathbf{x},\boldsymbol{\pi})\Big),\nonumber
\end{eqnarray}
where $I_s(\boldsymbol{\pi})=f\mathcal{P}_{-s}$, and $ \mathcal{P}_{-s} = \frac{1}{2} (1 -s \Lambda )$. Thus Eqs.~(\ref{MasterWigner}),(\ref{LRRf})-(\ref{LWFf}) provide a system of the Lindblad-Wigner-Dirac equations when the electron dynamics in the external field is quasiclassical.

\textit{Semiclassical Limit.} To bridge the quantum and classical regimes, we perform  FW transformation, which diagonalizes the free Dirac Hamiltonian. For simplicity, we apply FW transformation in the case of an electron in a time-independent magnetic field. After projecting out NES from the resulting equation we obtain the Wigner equation in the FW representation, see End Matter.
  
From the quasiclassical kinetic equation, we obtain the  corrected quasiclassical Ehrenfest equations employing the Wigner function $W_{FW}^+=\delta(\boldsymbol{\pi}-\boldsymbol{\pi}(t))\delta(\mathbf{x}-\mathbf{x}(t))\Phi(t)\Phi(t)^\dagger$, which in the case of the electron motion in a constant magnetic field reads
 \begin{eqnarray}
 \frac{d\boldsymbol{\pi}(t)}{dt}=\mathbf{F}+ \mathbf{f}_{\rm vac}- \mathbf{f}_{\rm rad}, \,\,\,
 \frac{d\mathbf{x}}{dt}=\mathbf{v}(t)+\delta \mathbf{v}_{\rm vac},\label{eq:final}
  \end{eqnarray}
where $\mathbf{F}$ is the Lorentz force, $\mathbf{f}_{\rm rad}=\sigma\frac{\gamma^2|m\gamma\ddot{\mathbf{x}}|^2}{2\pi m^2 }\mathbf{v}=\frac{d\varepsilon_{\rm rad}}{dt}\mathbf{v}$ the RR force. We obtain the physical value for the parameter of our model $\sigma=\frac{2\pi e^2}{3}$, imposing the requirement that the RR force  describes the momentum change due to the emitted radiation $\frac{d\varepsilon_{\rm rad}}{dt}=\tau_0\frac{\gamma^2\mathbf{F}^2}{m}$, with the characteristic RR time-parameter $\tau_0=\frac{2e^2}{3m}$.
The term $\mathbf{f}_{\rm vac}$ in Eq.~(\ref{eq:final}) is the contribution of VF  to the Lorentz force:
\begin{eqnarray}
\mathbf{f}_{\rm vac}=\frac{e}{c}(\delta \mathbf{v}_{\rm vac}\times \mathbf{B})+\frac{e}{c} [ \mathbf{v}\times ( \delta \mathbf{r}_{\rm vac}\cdot \boldsymbol{\nabla})\mathbf{B} ].\label{dF}
\end{eqnarray}
It is determined by the contributions of VF \textit{Zitterbewegung} to the  velocity $\delta \mathbf{v}_{\rm vac}=\frac{\sigma_-\lambdabar}{2m^2 \gamma}\mathbf{F}$, and to the coordinate $\delta \mathbf{r}= \frac{\sigma_-\lambdabar}{2m^2 \gamma} m\gamma \mathbf{v}$. 
We determine the second parameter of our model $\sigma_-= \frac{2e^2 }{3\lambdabar}$ requiring that Eq.~(\ref{eq:final}) leads to the LL equation (at the first-order in  $\tau_0 $):
\begin{eqnarray}
m\gamma \frac{d^2\mathbf{x}(t)}{dt^2}=\mathbf{F}-\frac{\tau_0\gamma^2\mathbf{F}^2}{m}
\mathbf{v} 
+ e\tau_0\left[\frac{\mathbf{F}\times \mathbf{B}}{m}
+\gamma\mathbf{v} \times\left(\mathbf{v}\cdot\nabla \right)\mathbf{B}\right].\nonumber \label{LL}
\end{eqnarray}
 While the interpretation of  the first $\sim\tau_0$ as energy dissipation RR term in LL equation is well known, the second $\sim\tau_0$ term  in Eq.~(\ref{LL}) represents the signature of VF in the classical dynamics, when the Zitterbewegung modifies the Lorentz force via Eq.~(\ref{dF}), inducing fluctuations 
\begin{eqnarray}
\delta \mathbf{v}_{\rm vac}=\frac{\tau_0}{2m}\mathbf{F}, \,\,\,\,\delta \mathbf{r}_{\rm vac}=\frac{\tau_0  }{2}\gamma \mathbf{v}.
\end{eqnarray}
The velocity fluctuation, which results in noncollienarity of the current and momentum, see $\delta \mathbf{v}_{\rm vac}$ term in Eq.~(\ref{eq:final}), has been already pointed out by Sokolov~\cite{sokolov2009renormalization}, however neglecting the second  VF contribution $\delta \mathbf{r}_{\rm vac}$ in Eq.~(\ref{dF}) to the Lorentz force modification, which is necessary to explicitly arrive to the LL equation.
  
VF contribute also in the  classical kinetic equation. The latter is derived expanding the Wigner equation in FW representation  up  to  $\hbar^2$ (as the parameters $\sigma_\pm$ contain $\hbar$  in the denominator):
\begin{eqnarray}
\label{kin_eq}
&&\frac{\partial f_0}{\partial t}
+ (\mathbf{v}+\delta \mathbf{v}_{\rm vac})\cdot\nabla f_0 +\nabla_{\boldsymbol{\pi}}\cdot(\mathbf{F}+\mathbf{f}_{\rm vac}+\mathbf{f}_{\rm rad})f_0   \nonumber\\
&=&\frac{\gamma\tau_0}{2}\left[ (\mathbf{F}\cdot\nabla_{\boldsymbol{\pi}})^2+ (\mathbf{v}.\nabla)^2
+ 2\mathbf{F}\cdot\{(\mathbf{v}\cdot\nabla)\nabla_{\boldsymbol{\pi}}\}\right]f_0. 
\end{eqnarray}
This equation includes  VF corrections to the velocity $\delta \mathbf{v}_{\rm vac}$ and coordinate $\delta \mathbf{r}_{\rm vac}$ discussed below for the classical equation of motion, and additionally the  corrections in the right side, which describe the VF induced classical diffusion in the coordinate ($\propto \nabla^2$ terms) as well as in the momentum space ($\propto\nabla^2_{\boldsymbol{\pi}}$ terms) and their cross-correlation ($\propto\nabla_{\boldsymbol{\pi}}\nabla$ terms). The possibility for the observation of the VF induced classical diffusion is discussed in the End Matter.

\textit{Conclusions.} We have developed a relativistic Lindblad model for the radiating electron in external fields  by incorporating both RR and VF effects into the Dirac equation via the ODM concept. The stabilization mechanism introduced through particle--antiparticle exchange  due to VFs is critical for preventing runaway solutions that could arise from the unbounded negative energy continuum. We constructed the phase space representation obtaining a compact expression for the Wigner equations in the case of the electron's quasiclassical dynamcs in the background fields. Performing a Foldy--Wouthuysen transformation, we derived corrected Ehrenfest equations that seamlessly bridge quantum and classical dynamics. The preservation of symmetry between positive- and negative-energy states, together with stabilization via particle--antiparticle exchange, confirms the robustness of our approach. In the classical limit, we confirm that that Zitterbewegung due to VFs renders electron velocity  to be not collinear with momentum, and induces  modifications in the Lorentz force, the traces of which are evident in the LL equation. We point out also the VF induced diffusion terms in the classical kinetic equation. 

Our framework opens several avenues for future work. In particular, a numerical investigation of the full Lindblad dynamics in external fields, as well as applications to multiparticle systems and strong-field QED, are promising directions. Furthermore, the connection between Zitterbewegung-induced noncommutativity and potential experimental signatures warrants deeper exploration.

 A.G.C acknowledges insightful discussions with Prof. Denys I. Bondar, Renan Cabrera and Prof. Herschel Rabitz during the early stages of this work \cite{campos2015excess}.

\section{End Matter}

{\bf{Remedy of the Lindbad model instability.}} In contrast to nonrelativistic quantum open systems, the relativistic generalization adds a new degree of freedom, the possibility of populating NES.
In the Dirac picture, an  electron undergoing RR energy dissipation 
dives into the negative energy continuum, and the system never equilibrates.  This problem is solved introducing the Lindblad operators that model PES/NES exchange.
 
To distinguish between PES and NES, a unitarily FW transformation is employed  $\hat{U}_{FW}=e^{i\hat{S}},\quad \hat{S}=-\frac{1}{2}\tan^{-1}\left(\frac{i\gamma^k\hat{\boldsymbol{p}}_k}{mc}\right)$, which diagonalizes the Hamiltonian $\hat{U}_{FW}\hat{H}_0\hat{U}_{FW}^\dagger=\gamma^0c\sqrt{\hat{\mathbf{p}}^2+m^2c^2}$, and the Dirac spinor  is represented as $\psi = \psi^+ + \psi^-$, with PES/NES energy components $\psi^+/\psi^-$ only. 

Usually, the open system description of the generalized amplitude damping channel \cite{khatri2020information,Nielsen} transitions between ``ground state" and the ``excited state" are introduced. Here, we associate the PES with a ``ground state" and the NES with an ``excited state", 
and introduce operators describing the transitions between them   
$$
E_1=\sqrt{\theta}\begin{pmatrix}
	 0 & 1 \\
	0 & 0
	\end{pmatrix},\quad  E_2=\sqrt{1-\theta}\begin{pmatrix}
	 0 & 0 \\
	1 & 0
	\end{pmatrix}
$$
where $\sqrt{\theta}$ is the probability of  the transition. If we call the open system operators $\hat{V}_\pm$ respectively, they should have the following forms in the FW representation
\begin{align}
	\hat{U}_{FW}\hat{V}_{+}\hat{U}^\dagger_{FW}&=if(\hat{\mathbf{p}})\,\gamma_5\gamma_0(1-\gamma_0)/2
	=f(\hat{\mathbf{p}})\begin{pmatrix}
	\boldsymbol{0} & \boldsymbol{1}\\
	\boldsymbol{0} & \boldsymbol{0} \nonumber
	\end{pmatrix},\\
	\hat{U}_{FW}\hat{V}_{-}\hat{U}^\dagger_{FW}&=if(\hat{\mathbf{p}})\,\gamma_0\gamma_5(1+\gamma_0)/2
	=f(\hat{\mathbf{p}})\begin{pmatrix}
	\boldsymbol{0} & \boldsymbol{0} \\
	\boldsymbol{1} & \boldsymbol{0}  
	\end{pmatrix}
	\label{Creation_annihilation}
\end{align}
where $\gamma_5=\gamma_0\gamma_1\gamma_2\gamma_3$, $f(\hat{\mathbf{p}})=\sqrt{2}\sqrt{\hat{\mathbf{p}}^2+m^2c^2}/mc$ which ensures that the Lindblad operators $\hat{V}_\pm$ resemble creation/annihilation operators connecting states with energy difference $\omega=2c^2(\mathbf{p}^2+m^2c^2)$ and $\boldsymbol{0}$, $\boldsymbol{1}$ are the $2\times2$ zero and identity matrices, respectively. Thus, the operators in (\ref{Creation_annihilation}) describe transitions from NES to PES upon the absorption of a virtual photon and vice versa.
For  the free electron  $ \hat{V}_{\pm}=\pm if\gamma_5\gamma_0\hat{\mathcal{P}}_\mp$, with the corresponding projector operators for PES (NES) $\hat{\mathcal{P}}_\pm =\frac{1}{2}\left(1\pm\hat{\Lambda}_0\right)$, and  the sign operator $\hat{\Lambda}_0=\hat{H}_0(\hat{H}_0^2)^{-1/2}$  with eigenvalues $+(-)$ for PES (NES) ($ \hat{\mathcal{P}}_\pm^2= \hat{\mathcal{P}}_\pm$ and $[ \hat{\mathcal{P}}_\pm,\hat{H}_0]=0$).
The sequence of transitions PES $\rightarrow$ NES $\rightarrow$ PES during the interaction mimics VF and induces  Zitterbewegung \cite{milonni2013quantum}.

\textbf{FW transformation.} To bridge the quantum and classical regimes, we perform the FW transformation, which diagonalizes the free Dirac Hamiltonian.
For simplicity, we consider the case 
of an electron in a time-independent magnetic field. After the FW transformation, the Hamiltonian in phase space to first order in the field strength becomes 
\begin{eqnarray}
\op{H}_{\mathrm{FW}} = \gam{0}\sqrt{\op{\boldsymbol{\pi}}^2+m^2}
-\frac{e}{2}\frac{\gam{0}\bm{\Sigma}\cdot \bvec{B}(\op{\mathbf{x}})}{\sqrt{\op{\boldsymbol{\pi}}^2+m^2}},
\label{eq:fw_hamiltonian_full2}
\end{eqnarray}
and the projected Wigner function, defined as $W^+_{\mathrm{FW}}=\op{P}_+W_{\mathrm{FW}}\op{P}_+$, satisfies an evolution equation obtained by substituting the FW-transformed operators into Eq.~\eqref{EqMoyalStar}. A detailed expansion yields $\mathbf{P}_+[ \gamma_0\sqrt{\boldsymbol{\pi}^2+m^2}, W_{FW}]\mathbf{P}_+=0 $.
After projecting out NES from the resulting equation we have
\begin{align}\label{PESEq}
	&\mathbf{P}_{+}\frac{\partial}{\partial t}W_{FW}\mathbf{P}_{+}+\mathbf{P}_{+}\frac{1}{2}\{\gamma^0e\mathbf{F}\cdot\nabla_{\boldsymbol{\pi}},W_{FW}\}\mathbf{P}_{+}\nonumber\\
	&+ \mathbf{P}_{+}\frac{1}{2}\{\gamma^0\mathbf{v}\cdot\nabla,W_{FW}\}\mathbf{P}_{+}-i\frac{e\mathbf{P}_{+}[\gamma^0\boldsymbol{\Sigma}\cdot\mathbf{B}(\mathbf{x}),W_{FW}]\mathbf{P}_{+}}{2\sqrt{\boldsymbol{\pi}^2+m^2}}\nonumber\\
	&= \frac{1}{\hbar}\left(\mathcal{L}_{RR}^{FW}(W_{FW}^+)
	+\widetilde{\mathcal{L}}_{VF}^{FW}(W_{FW}^+)\right) ,
\end{align}
where $\widetilde{\mathcal{L}}_{VF}^{FW}(W_{FW}^+)$ corresponds to the quantum corrections due to VF. 
The expression for $\mathcal{L}_{RR}^{FW}(W_{FW}^+)$ is derived using the operator Eq.~(\ref{eq:HjumpFW}) (the change $\tau\rightarrow-\tau$ is used):
\begin{align}\label{RRSemiCl}
\mathcal{L}_{RR}^{FW}(W_{FW}^+)&=\frac{\sigma}{2}\sum_{\eta=\pm1}\int_{-\infty}^\infty d\tau\int\frac{d^3\mathbf{k}}{(2\pi)^3\omega}
\mathcal{T}^{FW},
\end{align}
where $\mathcal{T}^{FW}=R_1 L W_{FW}^+(x,\boldsymbol{\pi}-\eta\mathbf{k}/2)R_2^\dagger
-\frac{L}{2}\{W_{FW}^+,R_2^\dagger R_1 \}$,
and $L$ is given in Eq.~(\ref{LLL}).

\begin{widetext}
The VF operators appearing in $\tilde{\mathcal{L}}_{VF}^{FW}$ in the FW representation are $V_{\pm}^{FW}=\pm i\gamma_5\gamma_0\frac{\sqrt{2}\boldsymbol{\pi}_0}{mc}\,(1\mp\gamma_0)/2$.
Thus, we have
\begin{align}\label{FWVFPS}
\tilde{\mathcal{L}}_{VF}^{FW}(W_{FW}^+)
&=\sum_{s=\pm}\frac{\sigma_{s}}{2}\Big(2i\gamma_5\gamma_0(F\star\mathbf{P}_{-s} W_{FW}^+\mathbf{P}_{-s}\star F)\,i\gamma_0\gamma_5
-F\star F\star \mathbf{P}_{-s}W_{FW}^+-W_{FW}^+\mathbf{P}_{-s}\star F\star F\Big),
\end{align}
where $F=\sqrt{2}\boldsymbol{\pi}_0/mc$. Using the properties of the projectors leads to
\begin{align}\label{FWVFfinal}
\tilde{\mathcal{L}}_{VF}^{FW}(W_{FW}^+)&=\frac{\sigma_{-}}{2}\Big(2i\gamma_5\gamma_0F\star W_{FW}^+\star F\,i\gamma_0\gamma_5
-(F\star F\star W_{FW}^++W_{FW}^+\star F\star F)\Big).
\end{align}
Therefore, to second order in $\hbar$ we have
\begin{align}
\frac{1}{\hbar}{\mathcal{L}}_{VF}^{FW}(W_{FW}^+)&=-\frac{\hbar^2\sigma_-}{2m^2c^6}\Bigg[e^2\boldsymbol{F}_i\boldsymbol{F}_j\frac{\partial^2}{\partial\boldsymbol{\pi}_i\partial\boldsymbol{\pi}_j}+\boldsymbol{v}_i\boldsymbol{v}_j\frac{\partial^2}{\partial\boldsymbol{x}_i\partial\boldsymbol{x}_j}+\frac{c(e\mathbf{F}\cdot\nabla)}{\boldsymbol{\pi}_0} \nonumber\\
&+\left[\frac{\mathbf{v}}{c}\times(\mathbf{v}\cdot\nabla e\mathbf{B})+\frac{e\mathbf{F}\times e\mathbf{B}}{\boldsymbol{\pi}_0}\right]\cdot(\nabla_{\boldsymbol{\pi}} )+2ce\mathbf{F}\cdot\{(\mathbf{v}\cdot\nabla)\nabla_{\boldsymbol{\pi}}\}\Bigg]\left(\gamma_5\gamma_0W_{FW}^+\gamma_0\gamma_5 + W_{FW}^+\right).
\end{align}
This procedure enables us to extract the kinetic equation in the quasiclassical limit.

\textbf{Quasiclassical Wigner kinetic equation.} 
The RR term can be simplified using the Fourier shift theorem:
\begin{eqnarray*}
&B_{\eta,I}(t_2,k)\star W\star B_{\eta,I}(t_1,k)^\dagger=\frac{\sigma}{2}e^{-i\eta\omega\tau}(\boldsymbol{\alpha}_2\cdot\boldsymbol{e}_\lambda)e^{\frac{iH\tau}{2}}e^{\frac{-iH'\tau}{2}}\star W(x,\boldsymbol{\pi}')\star e^{\frac{-iH'\tau}{2}}e^{\frac{iH\tau}{2}}(\boldsymbol{\alpha}_1\cdot\boldsymbol{e}_\lambda^\ast),\\
&\{W, B_{\eta,I}(t_1,k)^\dagger\star B_{\eta,I}(t_2,k)\}_\star=\frac{e^{-i\eta\omega\tau}\sigma}{2} 
 \{W,(\boldsymbol{\alpha}_1\cdot\boldsymbol{e}_\lambda^\ast)\star e^{iH\tau}e^{-iH'\tau}(\boldsymbol{\alpha}_2\cdot\boldsymbol{e}_\lambda)\}_\star,
\end{eqnarray*}
where $\{A,B\}_\star=A\star B+B\star A$, $\boldsymbol{\pi}'=\boldsymbol{\pi}-\eta\mathbf{k}$ and $H'=H(\boldsymbol{\pi}')$.
Summing over photon polarization leads to
\begin{align*}
&\mathcal{L}_{RR}(W)=\frac{\sigma}{2}\sum_{\eta=\pm1}\int_{-\infty}^\infty d\tau\int\frac{d^3\mathbf{k}}{(2\pi)^3\omega}\Big[P_{ij}(\mathbf n)\alpha_{2}^i L(\tau/2)\star 
  W(x,\boldsymbol{\pi}-\eta\mathbf{k}) \star L(-\tau/2)^\dagger\alpha_{1}^j
	 -\frac{P_{ij}(\mathbf n)}{2}\{W,\alpha_1^i \stackon{\star}{B} L(\tau)\alpha_2^j\}_\star\Big],
\end{align*}
where $L(\tau)=e^{i(H-\eta\omega)\tau}e^{-iH'\tau}$. Like we did with the VF term, let us separate the effect of the star product term by term. The easiest 
is the anticommutator, which is
\begin{align}\label{AC_RR}
&(\alpha_1^i \stackon{\star}{B} L(\tau)\alpha_2^j) \stackon{\star}{B}\overleftarrow{W}\Big(t,\mathbf{x}+\frac{i\hbar\nabla_{\boldsymbol{\pi}}}{2},\boldsymbol{\pi}-\frac{i\hbar\nabla_{\mathbf{x}}}{2}\Big) 
 +\overrightarrow{W}\Big(t,\mathbf{x}-\frac{i\hbar\nabla_{\boldsymbol{\pi}}}{2},\boldsymbol{\pi}+\frac{i\hbar\nabla_{\mathbf{x}}}{2}\Big)\stackon{\star}{B}(\alpha_1^i \stackon{\star}{B} L(\tau)\alpha_2^j)
\end{align}
while the first term is $\big(\alpha_{2}^i L(\tau/2)\stackon{\star}{B} \overleftarrow{W}\Big(t,\mathbf{x}+\frac{i\hbar\nabla_{\boldsymbol{\pi}}}{2},\boldsymbol{\pi}'\Big)\big) \stackon{\star}{B} \overleftarrow{L(-\tau/2)^\dagger\alpha_{1}^j}\Big(\boldsymbol{\pi}-\frac{i\hbar\nabla_{\mathbf{x}}}{2}\Big)$.
\end{widetext}
Let us now disentangle the product in $L$. Taking the $\tau$ derivative gives
\begin{align*}
\partial_\tau L(\tau)&=ie^{i(H-\eta\omega)\tau}(H-\eta\omega-H')e^{-iH'\tau}\\& 
 =i \eta(-\omega+\mathbf{k}\cdot\boldsymbol{\alpha}'(t_2))L(\tau).
\end{align*}

Hence, we get
\begin{align*}
L(\tau)=\mathcal{T}\exp\left(-i\eta\int_0^\tau d\tau'\left[\omega-\mathbf{k}\cdot\boldsymbol{\alpha}'(t_1+\tau')\right]\right)L(0),
\end{align*}
where $\mathcal{T}$ is the time-ordering operator. Noting that $d\mathbf{x}'(t)/dt=\boldsymbol{\alpha}'(t)$, where $\mathbf{x}'(t)=e^{iH't}\mathbf{x}e^{-iH't}$ we have
\begin{align}\label{Lexpression}
L(\tau)=\mathcal{T}\exp\left(-i\eta\left[\omega\tau-\mathbf{k}\cdot(\mathbf{x}'(t_2)-\mathbf{x}'(t_1))\right]\right)L(0).
\end{align}
Using an approximation with omitting $\mathcal{T}$  for $L(\tau)$ yields
\begin{align}\label{Ldissentangled}
L(\tau)=\exp\left(-i\eta\left[\omega\tau-\mathbf{k}\cdot(\mathbf{x}'(t_2)-\mathbf{x}'(t_1))\right]\right)L(0).
\end{align}

\textbf{Low recoil limit.}  Let us consider the low recoil limit, $\boldsymbol{\pi}_0'\approx\boldsymbol{\pi}_0$. Performing the $\mathbf{k}$ integration and summing over $\eta$ in Eq.~(\ref{EhrenfestRR}), one obtains
\begin{align*}
\mathcal{L}_{RR}^\dagger(\hat{\boldsymbol{\pi}})&=\int_{-\infty}^{\infty}\frac{\sigma|\dot{\mathbf{v}}|^2\tau^2 d\tau   }{2\pi^2\left(\tau ^2-\mathbf{r}^2\right)^2}i\mathbf{r}
\end{align*}
where $\mathbf{r}=\mathbf{x}(t_2)-\mathbf{x}(t_1)$. We expand over $\tau$ assuming LCFA, when the photon emission formation length is small  in the  typical scale of the external field, and obtain $\mathbf{r}\approx\tau\dot{\mathbf{x}}+\frac{\tau^3\ddot{\mathbf{v}}}{24}$ and $\dddot{\mathbf{x}}\cdot\dot{\mathbf{x}}=-\ddot{\mathbf{x}}^2$,   
$\mathbf{r}^2- \tau^2\approx- (\tau^2/\gamma^2) (1+ \tau^2\gamma^2\ddot{\mathbf{x}}^2/12)$ \cite{Baier_b_1994}, which finally yields  
\begin{align}\label{eq:LRRp}
\mathcal{L}_{RR}^\dagger(\hat{\mathbf{p}})\approx-\frac{\sigma  \gamma ^4 |\ddot{\mathbf{x}}|^2}{2\pi  }\dot{\mathbf{x}}.
\end{align}
Thus, we see that  the Lindbad operators of Eq.~(\ref{eq:Hjump}) in the limit of small recoil induces the friction term $\mathcal{L}_{RR}^\dagger(\hat{\boldsymbol{\pi}}(t))$ proportional to velocity, i.e., $\dot{\mathbf{x}}$ in the Ehrenfest momentum equation, fulfilling our aim. 

 \textbf{Classical vs quantum diffusion.}  It is known that the quantum corrections to the Boltzmann equation due to RR induce diffusion \cite{neitz2013stochasticity}. We can estimate the ratio of the quantum diffusion coefficient in the momentum space to the classical one from Eq.~(\ref{kin_eq}): $D_q/D_c\sim \gamma^3 E/E_{cr}$, where $E$ is the background field strength, and  $E_{cr} =m^2/e$ is the QED critical field. Note that the ratio of the diffusion term in the right side of Eq.~(\ref{kin_eq}) to that of the RR term in the left side of Eq.~(\ref{kin_eq}) is ${\rm RR/Diff}\sim m/(\gamma\Delta p)$, with the momentum spread $\Delta p$. Thus, for the observation of the classical diffusion vs the effect of RR force, one needs not large $\gamma$. The classical diffusion could have comparable effect to the quantum one  at rather small $E/E_{cr}$, but the ratio of the RR force to the background field is $f_{\rm rad}/F\sim \alpha \gamma^2 E/E_{cr} $. Trading off the conditions above, we may point out the optimal conditions for the observation of the classical diffusion: at feasible $E/E_{cr}\sim 10^{-3}$ \cite{Yoon_2021} and $\gamma\sim 10$, one has comparable quantum and classical diffusion $D_q/D_c\sim 1$, with still not negligible effect of RR: $f_{\rm rad}/F\sim 10^{-3}$ and ${\rm RR/Diff}\sim 10^{-1}$ ($\Delta p \sim m$).

\begin{widetext}
\section{ Supplemental material for ``Relativistic Lindblad description of the electron radiative dynamics''}

\subsection{The dynamical equation from Choi-Kraus' theorem}\label{DerivingLindblad}
Here we propose the relativistic generalization of the radiation reaction Lindblad term used in the main text for a general Lindblad operator.

According to Choi-Kraus' theorem (see, for instance, Ref. \cite{manzano2020short}) a linear time evolution of the density matrix $\hat{P}(t)=\sum_k\hat{T}_k(t,t_0)\hat{P}(t_0)\hat{T}(t,t_0)_k^\dagger$ for some evolution operators $\hat{V}_k$ not necessarily Hermitian is CPT if and only if $\sum_k\hat{T}_k^\dagger\hat{T}_k=1$. Let us represent the evolution equation in the following form
\begin{align}\label{nonUnitaryLindblad}
\hat{P}(t)-\hat{P}(t_0)&=\frac{i}{\hbar}\int_{t_0}^{t}dt'\left[\hat{P}(t_0),\hat{H}(t')\right]\nonumber\\
&+\sum_{k} \left(\hat{T}_k\hat{P}\hat{T}_k^\dagger-\frac{1}{2}\left\{\hat{T}_k^\dagger \hat{T}_k,\hat{P}\right\}\right).
\end{align}
Considering
\begin{align*}
 \hat{T}_1(t)&=\int_{t_0}^t dt' \hat{B}(t'),
\end{align*}
the explicit form of $ \hat{B}$ is yet to be given. 
Since $\sum_k\hat{T}_k\hat{P}(t_0)\hat{T}_k^\dagger=\hat{P}(t)$, Eq. (\ref{nonUnitaryLindblad}) becomes 
\begin{align}\label{nonUnitaryLindblad2}
&\hat{P}(t)=\hat{P}(t_0)+\frac{i}{\hbar}\int_{t_0}^{t}dt'\left[\hat{P}(t_0),\hat{H}(t')\right]\nonumber\\
&+ \int_{t_0}^{t}dt_1\int_{t_0}^{t}dt_2\Big( \hat{B}(t_2)\hat{P}(t_0) \hat{B}(t_1)^\dagger\nonumber\\
&-\frac{1}{2}\left\{ \hat{B}(t_1)^\dagger \hat{B}(t_2),\hat{P}(t_0)\right\}\Big).
\end{align}
In the third term on the right hand side make the replacements $t'=(t_1+t_2)/2$ and $\tau=t_2-t_1$. For a sufficiently extended time interval $t$ one can extend the limits of the relative time $\tau$ to $\pm\infty$, leading to
\begin{align}\label{nonUnitaryLindblad2a}
&\hat{P}(t)=\hat{P}(t_0)+\frac{i}{\hbar}\int_{t_0}^{t}dt'\left[\hat{P}(t_0),\hat{H}(t')\right]\nonumber\\
&+ \frac{1}{2}\int_{t_0}^{t}dt'\int_{-\infty}^{\infty}d\tau\Big( \hat{B}(t'+\tau/2)\hat{P}(t_0) \hat{B}(t'-\tau/2)^\dagger\nonumber\\
&-\frac{1}{2}\left\{ \hat{B}(t'-\tau/2)^\dagger \hat{B}(t'+\tau/2),\hat{P}(t_0)\right\}\Big).
\end{align}
After making the following substitutions $t_0\rightarrow t$, $t\rightarrow t+\delta t$, applying the trapezoidal rule for $\delta t\ll1$,
$\int_t^{t+\delta t}dt'f(t')=\delta tf(t+\delta t/2)+\mathcal{O}(\delta t^3)$
and taking the limit $\delta t\rightarrow0$ we finally arrive at
\begin{align}\label{nonUnitaryLindblad3}
&\frac{\partial\hat{P}(t)}{\partial t}=\frac{i}{\hbar}\left[\hat{P}(t),\hat{H}(t)\right]\nonumber\\
&+ \frac{1}{2}\int_{-\infty}^{\infty}d\tau\Big( \hat{B}(t+\tau/2)\hat{P}(t) \hat{B}(t-\tau/2)^\dagger\nonumber\\
&-\frac{1}{2}\left\{ \hat{B}(t-\tau/2)^\dagger \hat{B}(t+\tau/2),\hat{P}(t)\right\}\Big).
\end{align}
Hereinafter $\{.,.\}$ and $[.,.]$ will stand for the anticommutator and commutator, respectively. 

\subsection{Renormalization as open system}

The idea presented in the previous section can also be applied to the renormalization procedure in QFT.

Assume the system's Hamiltonian is of the form
\begin{align}
	\hat{H} = \hat{H}_0 + \lambda_0 \hat{V}. 
\end{align}
We shall refer to it as the bare Hamiltonian because it contains a `bare' (i.e., not directly observed) value of the parameter $\lambda_0$. For the Dirac equation within QED, $\lambda_0$ stands for the bare mass or bare charge of an electron. According to the renormalization theory, an experimentally observed parameter is $\lambda = \lambda _0+ \delta \lambda$, where $\delta \lambda$ is \emph{divergent} contribution coming from the particle interaction with vacuum. Following Dyson \cite{dyson1949radiation}, we write
\begin{align}
	\hat{H} = \hat{H}_0 + (\lambda_0 + \delta \lambda) \hat{V}  - \delta \lambda \hat{V}  
		= \hat{H}_0 + \lambda \hat{V} - \delta \lambda \hat{V}.
\end{align}
Equation of motion for the propagator is
\begin{align}
	i\hbar \frac{\partial}{\partial t} \hat{U}(t, t') = \left[ \hat{H}_0 + \lambda \hat{V} - \delta \lambda \hat{V} \right] \hat{U}(t, t').
\end{align}
We go to the interaction picture by representing this propagator as
\begin{align}
	\hat{U}(t, t') = e^{-it (\hat{H}_0 + \lambda \hat{V}) / \hbar} \hat{U}_I(t, t'),
\end{align}
hence
\begin{align}\label{EqPropagatorInterPicture}
	i\hbar  \frac{\partial}{\partial t} \hat{U}_I(t, t') &= -\delta \lambda \hat{V}_I(t) \hat{U}_I(t, t'), \\
	\hat{V}_I(t) &= e^{it (\hat{H}_0 + \lambda \hat{V}) / \hbar} \hat{V} e^{-it (\hat{H}_0 + \lambda \hat{V}) / \hbar}.
\end{align}
The time-ordered exponent provides the solution to Eq. \eqref{EqPropagatorInterPicture} 
\begin{align}
	\hat{U}_I(t, t') &= \hat{\mathcal{T}} \exp\left[ \frac{-i}{\hbar} \int_{t'}^{t} (-\delta \lambda) \hat{V}_I(\tau) d\tau \right] \notag\\
		&= 1 - i \left[ \hat{T}_1 + \hat{T_2} + O\left( \delta \lambda^3 \right) \right], \label{EqTimeOrderedExp} \\
	\hat{T}_1 &= \frac{-\delta \lambda}{\hbar} \int_{t'}^{t} d\tau_1 \, \hat{V}_I(\tau_1), \notag\\
	\hat{T}_2 &= \frac{-i\delta \lambda^2}{\hbar^2} \int_{t'}^{t} d\tau_1 \int_{t'}^{\tau_1} d\tau_2 \, \hat{V}_I(\tau_1) \hat{V}_I(\tau_2). \notag
\end{align}
By grouping terms of the same order in $\delta \lambda$, the unitary condition $\hat{U}_I \hat{U}_I^{\dagger} = 1$ implies that
\begin{align}\label{EqConditionsOnT1T2}
	\hat{T}_1^{\dagger} = \hat{T}_1, \qquad i(\hat{T}_2 - \hat{T}_2^{\dagger}) = \hat{T}_1 \hat{T}_1^{\dagger}.
\end{align}
Using \eqref{EqTimeOrderedExp} and \eqref{EqConditionsOnT1T2}, we obtain 
\begin{align}\label{EqBaier}
	 \hat{U}_I \hat{\rho}_I \hat{U}_I^{\dagger} -  \hat{\rho}_I = &-i \Big[ \hat{T}_1 + \frac{\hat{T}_2 + \hat{T}_2^{\dagger}}2, \hat{\rho}_I \Big] \notag\\
	 	& + \hat{T}_1 \hat{\rho}_I \hat{T}_1 - \frac{1}{2} \{\hat{T}_1 \hat{T}_1, \hat{\rho}_I \} + O\left( \delta \lambda^3 \right).
\end{align}
Such a representation of the perturbative expansion was originally reported in Eq. (3.6) from \cite{bauier1972radiative}. Substituting $t \to t + \delta t$ and $t' \to t$ into \eqref{EqTimeOrderedExp} and utilizing the trapezoidal and rectangular integration formulae 
\begin{align}
	\int_{t}^{t+\delta t} f(\tau) d\tau &= f\left(t + \frac{\delta t}{2}\right) \delta t + O\left( \delta t^3 \right) \notag\\
		&= f\left(t + \delta t\right) \delta t + O\left( \delta t^2 \right),
\end{align}
one obtains the following estimates
\begin{align}
	\hat{T}_1 &= -\delta t \delta \lambda \, \hat{V}_I( t + \delta t / 2) / \hbar + O\left( \delta t^3 \right), \label{EqT1Asymptotic} \\
	\hat{T}_2 &= -i \left[ \delta t \delta \lambda \, \hat{V}_I( t + \delta t / 2) / \hbar\right]^2 +  O\left( \delta t^3 \right). \label{EqT2Asymptotic}
\end{align}
Substituting Eqs. \eqref{EqT1Asymptotic} and \eqref{EqT2Asymptotic} into Eq. \eqref{EqBaier} and taking the limit $\delta t \to 0$, $(\hat{U}_I \hat{\rho}_I \hat{U}_I^{\dagger} -  \hat{\rho}_I ) /\delta t \to \partial \hat{\rho}_I / \partial t$ and $\delta \lambda \to \infty$ such that $\gamma = \delta t \delta \lambda^2 /\hbar^2$ is fixed value, we obtain
\begin{align}
	\frac{\partial \hat{\rho}_I}{\partial t} =& \frac{i\delta \lambda}{\hbar} \left[ \hat{V}_I(t), \hat{\rho}_I \right] \notag\\
		& + \gamma \left( \hat{V}_I(t) \hat{\rho}_I \hat{V}_I(t) - \frac{1}{2}\left\{ \hat{V}_I(t)\hat{V}_I(t), \hat{\rho}_I \right\} \right).
\end{align}
Going back from the interaction picture to the Schr\"{o}dinger picture, 
$\hat{\rho} = e^{-it (\hat{H}_0 + \lambda \hat{V}) / \hbar} \hat{\rho}_I e^{it (\hat{H}_0 + \lambda \hat{V}) / \hbar}$,
we get
\begin{align}\label{EqGenericMasterEq}
	\frac{\partial \hat{\rho}}{\partial t} &= \frac{-i}{\hbar} \left[ \hat{H}_0 + \lambda_0 \hat{V} , \hat{\rho} \right] 
		 + \gamma \left( \hat{V} \hat{\rho}\hat{V} - \frac{1}{2}\left\{ \hat{V}\hat{V}, \hat{\rho} \right\} \right) \notag\\
	&= \frac{-i}{\hbar} \left[ \hat{H}_0 + \lambda_0 \hat{V} , \hat{\rho} \right]
		-\frac{\gamma}{2} \left[\hat{V}, \left[\hat{V}, \hat{\rho} \right] \right].
\end{align}
Therefore, the renormalization necessarily leads to the Lindblad master equation, where $\lambda_0$ and $\gamma$ are considered to be free parameters. It is interesting to notice that the markovian master equations in quantum theory emerge as approximation to the exact non-markovian dynamics. However, since the renormalization theory describes coupling to a bath with infinite bandwidth (i.e., arbitrary small spacial scales) the resulting interaction must necessarily be Markovian.

Similar to the master equations studied in Ref. \cite{vuglar2018nonconservative}, the master equation \eqref{EqGenericMasterEq} predicts the purity loss as
\begin{align}
	\frac{d}{dt} \Tr \left( \hat{\rho}^2 \right) = 2\gamma \Tr\left( (\hat{\rho}\hat{V})^2 - \hat{\rho}^2 \hat{V}^2 \right) \leq 0.
\end{align}

We note an Ehrenfest-like relation for the observable $\hat{O}$
\begin{align}
	\frac{d}{dt} \langle \hat{O} \rangle = \frac{-i}{\hbar} \left\langle [ \hat{O}, \hat{H} ] \right\rangle 
		-\frac{\gamma}{2} \left\langle [\hat{V}, [\hat{V}, \hat{O} ] ] \right\rangle.
\end{align}

The master equation \eqref{EqGenericMasterEq} is general and applicable also to the condense matter physics where the renormalization is utilized (e.g., see \cite{mattuck1992guide}).


\subsection{Quantum Ehrenfest equation for the kinetic momentum}

Here we provide a more general expression for the Ehrenfest relation $\hat{\boldsymbol{\pi}}$ without applying a FW transformation and projection as follows. To evaluate the dissipating term in the momentum Ehrenfest equation,
we calculate the expectation value $\langle\hat{O}\rangle=\frac{1}{4}\Tr [\hat{P}\hat{O} ]$ of a general observable $\hat{\mathcal{O}}$ related to the Ehrenfest relation which stems from the Lindbad dissipating operator: 
\begin{align}\label{EhrenfestRR}
\mathcal{L}_{RR}^\dagger(\hat{\mathcal{O}})&=\int d\tau\sum_{\eta=\pm1}\int\frac{d^3\boldsymbol{k}\sigma }{2(2\pi)^{3}\omega}
\sum_\lambda\mathcal{T},
\end{align}
where 
\begin{align*}
\mathcal{T}&=\hat B_{\eta,\lambda}\left(\frac{\tau}{2},k\right)^\dagger\left[\mathcal{O},\hat B_{\eta,\lambda}\left(-\frac{\tau}{2},k\right)\right]\\
&+\left[\hat B_{\eta,\lambda}\left(\frac{\tau}{2},k\right)^\dagger,\mathcal{O}\right]\hat B_{\eta,\lambda}\left(-\frac{\tau}{2},k\right).
\end{align*}
For $\hat{\mathcal{O}}=\hat{\boldsymbol{\pi}}$, we have
\begin{align*}
&\mathcal{T}=\frac{e^{-i\eta\omega\tau}}{2}U\left(\frac{\tau}{2}\right)\Bigg(\left[\hat B_{\eta,\lambda}\left(k\right)^\dagger,\boldsymbol{\alpha}(-\tau/2)\times e\mathbf{B}\right]U^\dagger(\tau)\hat B_{\eta,\lambda}\left(k\right)\\
&+\hat B_{\eta,\lambda}\left(k\right)^\dagger U^\dagger(\tau)\left[ \boldsymbol{\alpha}(\tau/2)\times e\mathbf{B},\hat B_{\eta,\lambda}\left(k\right)\right]\\
&-2\eta\mathbf{k}\,\boldsymbol\alpha\cdot\mathbf e_\lambda^\ast U'^\dagger(\tau)\boldsymbol\alpha\cdot\mathbf e_\lambda\Bigg)U\left(\frac{\tau}{2}\right)\\
&=e^{-i\eta\omega\tau}\Big(U\left(\frac{\tau}{2}\right)e^{i\eta\mathbf{k}\cdot\hat{\mathbf{x}}}\boldsymbol\alpha\cdot\mathbf e_\lambda^\ast U^\dagger(\tau)\left(\boldsymbol{\alpha}(\tau/2)\times e\mathbf{B}-\eta\mathbf{k}\right)\\
&\times\boldsymbol\alpha\cdot\mathbf e_\lambda e^{-i\eta\mathbf{k}\cdot\hat{\mathbf{x}}}U\left(\frac{\tau}{2}\right)\\
&-\frac{1}{2}\left\{\boldsymbol{\alpha}(\tau/2)\times e\mathbf{B},U\left(\frac{\tau}{2}\right)\boldsymbol\alpha\cdot\mathbf e_\lambda^\ast U'^\dagger(\tau)\boldsymbol\alpha\cdot\mathbf e_\lambda U\left(\frac{\tau}{2}\right)\right\}\Big)
\end{align*}
where $'$ means the replacement $\hat{\boldsymbol{\pi}}'=\hat{\boldsymbol{\pi}}-\eta\mathbf{k}$, $\boldsymbol{\alpha}(\tau)=\int_0^\tau ds e^{isH}\boldsymbol{\alpha}e^{-isH}$ and $\boldsymbol{\alpha}(-\tau)=\int_{-\tau}^0 ds e^{isH}\boldsymbol{\alpha}e^{-isH}$. One should note that $\boldsymbol{\alpha}(\tau/2)\times e\mathbf{B}=\int_0^\frac{\tau}{2} ds \boldsymbol{\alpha}(s)\times e\mathbf{B}=\int_0^\frac{\tau}{2} ds \dot{\boldsymbol{\pi}}(s)=\boldsymbol{\pi}(\tau/2)-\boldsymbol{\pi}(0)$. Hence,
\begin{align*}
&\mathcal{T}=e^{-i\eta\omega\tau}\Big(U\left(\frac{\tau}{2}\right)e^{i\eta\mathbf{k}\cdot\hat{\mathbf{x}}}\boldsymbol\alpha\cdot\mathbf e_\lambda^\ast U^\dagger(\tau)\left(\boldsymbol{\pi}(\tau/2)-\boldsymbol{\pi}(0)-\eta\mathbf{k}\right)\\
&\times\boldsymbol\alpha\cdot\mathbf e_\lambda e^{-i\eta\mathbf{k}\cdot\hat{\mathbf{x}}}U\left(\frac{\tau}{2}\right)\\
&-\frac{1}{2}\left\{\boldsymbol{\pi}(\tau/2)-\boldsymbol{\pi}(0),U\left(\frac{\tau}{2}\right)\boldsymbol\alpha\cdot\mathbf e_\lambda^\ast U'^\dagger(\tau)\boldsymbol\alpha\cdot\mathbf e_\lambda U\left(\frac{\tau}{2}\right)\right\}\Big)
\end{align*}

\subsection{The SemiClassical limit}\label{classical}

In this section, we closely follow the methodology of Refs. \cite{cabrera2016dirac,cabrera2019operational}. To first order in $\hbar$ the Moyal product becomes
\begin{align}\label{Moyal1st}
\star\approx1+\frac{i\hbar}{2}\left(\sum_{j=1}^3\left[ 
		\overleftarrow{\frac{\partial}{\partial \boldsymbol{x}^j}} \overrightarrow{\frac{\partial}{\partial \boldsymbol{\pi}^j}} -
		\overleftarrow{\frac{\partial}{\partial \boldsymbol{\pi}^j}} \overrightarrow{\frac{\partial}{\partial \boldsymbol{x}^j}}\right]
	+2\mathbf{L}_1 \right).
\end{align}
The upper arrows indicate the direction in which the derivatives act. With the above definition of the Moyal to lowest order in $\hbar$, the expression for $\hat{V}_+$ on the phase space can simply be taken as
\begin{align}\label{VPS1a}
V_{\pm}(\boldsymbol{\pi})&=\pm i\gamma_5\gamma_0f(\boldsymbol{\pi})\mathcal{P}_{\mp}\nonumber\\
&=\pm\frac{i\gamma_5\gamma_0}{\sqrt{2}}\left(\frac{\sqrt{\boldsymbol{\pi}^2+m^2c^2}}{mc}\mp\frac{H_A}{mc^2}\right),\nonumber\\
H_A&=-c\gamma^0\gamma^k\boldsymbol{\pi}_k+\gamma^0mc^2.
\end{align}
For each point in phase space $(\mathbf{x},\mathbf{p})$ the matrix valued function $H_A$ is Hermitian with the two double degenerate eigenvalues
$$
ch_{\pm}=\pm c\sqrt{\boldsymbol{\pi}^2+m^2c^2},
$$
which are identified with the classical Hamiltonians for particles ($+$) and antiparticles ($-$).
In this section, we will work in the FW representation. As demonstrated in Ref. \cite{eriksen1958foldy}, in the case of a magnetostatic field, an exact expression for $\hat{U}_{FW}$ exists and is given by
\begin{align}\label{UFW}
\hat{U}_{FW}=e^{i\hat{S}},\quad \hat{S}=-\frac{1}{2}\tan^{-1}\left(\frac{i\gamma^k\hat{\boldsymbol{\pi}}_k}{mc}\right).
\end{align}
 
\comm{leading to the closed form 
\begin{align}\label{UFWEx}
\hat{U}_{FW}&=\frac{(1+\gamma_0\hat{H}(\hat{H}^2)^{-1/2})}{\sqrt{2}}\left[1+\left(\frac{\hat{H}^2}{m^2c^4}\right)^{-1/2}\right]^{-1/2},\nonumber\\
\hat{U}_{FW}^\dagger&=\frac{(1+\hat{H}(\hat{H}^2)^{-1/2}\gamma_0)}{\sqrt{2}}\left[1+\left(\frac{\hat{H}^2}{m^2c^4}\right)^{-1/2}\right]^{-1/2}.
\end{align}
}
The transformed Hamiltonian is
\begin{align}\label{HFW}
\hat{H}_{FW}=c\gamma_0\sqrt{\hat{\boldsymbol{\pi}}^2+m^2c^2-\frac{e\hbar}{c}\boldsymbol{\Sigma}\cdot\mathbf{B}(\hat{\mathbf{x}})}.
\end{align}
To first order in $\hbar$, the FW Hamiltonian in the phase space becomes
\begin{align}\label{firstOHFW}
H_{FW}^1&=c\gamma_0\sqrt{\boldsymbol{\pi}^2+m^2c^2}-\frac{e\hbar}{2c\sqrt{\boldsymbol{\pi}^2+m^2c^2}}\gamma_0\boldsymbol{\Sigma}\cdot\mathbf{B}(\mathbf{x}),
\end{align}
which is the Hamiltonian we will use hereinafter.
\comm{
Let us now consider the operator appearing on the right hand side of Eq. (\ref{UFWEx}). 
It can be written as
\begin{align}\label{NegSqrtOp}
&\left[1+\left(\frac{\hat{H}^2}{m^2c^4}\right)^{-1/2}\right]^{-1/2}=\frac{1}{\sqrt{\pi}}\int_0^\infty duu^{-1/2}e^{-u}e^{-u\left(\frac{\hat{H}^2}{m^2c^4}\right)^{-1/2}},\,\left(\frac{\hat{H}^2}{m^2c^4}\right)^{-1/2}=\frac{1}{\sqrt{\pi}}\int_{0}^\infty duu^{1/2}e^{-u\frac{\hat{H}^2}{m^2c^4}},\\
&\frac{\hat{H}^2}{m^2c^4}=\frac{\hat{\boldsymbol{\pi}}^2+m^2c^2}{m^2c^2}-\frac{e\hbar}{m^2c^3}\mathbf{\Sigma}\cdot\mathbf{B}(\hat{\mathbf{x}}).
\end{align}
To first order in $\hbar$, we have
\begin{align*}
&\exp\left(-u\frac{\hat{H}^2}{m^2c^4}\right)=\exp\left(-u\left[\frac{\boldsymbol{\pi}^2+m^2c^2}{m^2c^2}\right]\right)\left\{1+u\frac{e\hbar}{m^2c^3}\mathbf{\Sigma}\cdot\mathbf{B}(\mathbf{x})\right\},\,\frac{i}{\sqrt{\pi}}\int_{-\infty}^0 duu^{-1/2}e^{u\frac{\hat{H}^2}{m^2c^4}}=\frac{mc}{\boldsymbol{\pi}_0}\left(1+\frac{e\hbar}{2\boldsymbol{\pi}_0^2c}\mathbf{\Sigma}\cdot\mathbf{B}(\mathbf{x})\right).\\
&\frac{1}{\sqrt{\pi}}\int_0^\infty duu^{-1/2}e^{-u}e^{-u\left(\frac{\hat{H}^2}{m^2c^4}\right)^{-1/2}}=\frac{1}{\sqrt{1+mc/\boldsymbol{\pi}_0}}\left(1-\frac{e\hbar m}{4\boldsymbol{\pi}_0^2(mc+\boldsymbol{\pi}_0)}\mathbf{\Sigma}\cdot\mathbf{B}\right).
\end{align*}
}
On the phase space, we consider the following expression for the FW transformation
\begin{align}
U_{FW}=\sqrt{\frac{\boldsymbol{\pi}_0}{2(mc+\boldsymbol{\pi}_0)}}\left(1+\frac{\vec{\gamma}\cdot\boldsymbol{\pi}+mc}{\boldsymbol{\pi}_0}\right)\label{FWCl2}.
\end{align}
Note that (\ref{FWCl2}) are matrix valued functions on the classical phase space.
It is very important to note the following identity
\begin{align}\label{classicalspinor}
U^\dagger_{FW}\psi&=U^\dagger_{FW}\begin{pmatrix}
	\Phi\\
	\mathbf{0}
	\end{pmatrix}=\sqrt{\frac{\boldsymbol{\pi}_0+mc}{2\boldsymbol{\pi}_0}}\begin{pmatrix}
	\Phi\\
	\frac{\vec{\sigma}\cdot\boldsymbol{\pi}}{\boldsymbol{\pi}_0+mc}\Phi\end{pmatrix}\nonumber\\
	&=\frac{u_{\boldsymbol{\pi}}}{\sqrt{2\boldsymbol{\pi}_0}}.
\end{align}
where $u_{\boldsymbol{\pi}}$ is the Dirac $4$-spinor, $\Phi=(\Phi_1,\Phi_2)^T$, and $\mathbf{0}=(0,0)^T$. From this follows  Baier's photon emission probability in the quasi-classical limit.
\end{widetext}
\bibliography{bib-relativity-2}

\end{document}